\documentclass[12pt]{article}
\usepackage{graphicx}
\usepackage{amsmath,amssymb}
\newcommand{\ket}[1]{\left| #1 \right>}
\newcommand{\braket}[1]{\left<#1\right>}
\newcommand{\id}{{\cal I}}

\renewcommand{\thepage}{}
\makeatletter
\@addtoreset{equation}{section}
\renewcommand{\theequation}{\thesection.\@arabic\c@equation}
\makeatother
\renewcommand{\thefootnote}{\fnsymbol{footnote}}
\begin{document}
\begin{titlepage}
\title{
\vspace*{-4ex}
\hfill
\begin{minipage}{3.5cm}
\normalsize YITP-05-42  \\
\normalsize hep-th/0508196
\end{minipage}\\
\vspace{4ex}
\bf 
Twist Symmetry and Classical Solutions
in Open String Field Theory
\vspace{5ex}}
\author{Syoji Zeze$^1$\footnote{E-mail address:
zeze@yukawa.kyoto-u.ac.jp}  
\vspace{2ex}\\
$^1${\it Yukawa Institute for Theoretical Physics,}\\
{\it Kyoto University, Japan}\\
}
\date{August, 2005}
\maketitle
\vspace{7ex}

\begin{abstract}
\normalsize
\baselineskip=19pt plus 0.2pt minus 0.1pt
We construct classical solutions of open
string field theory which are not invariant under
ordinary twist operation.
From detailed analysis of the moduli space of the solutions,
it turns out that our solutions become nontrivial at
boundaries of the moduli space.
The cohomology of the modified BRST operator
and the CSFT potential evaluated by the level truncation method
strongly support the fact that our nontrivial solutions 
correspond to the closed string vacuum.
We show that the nontrivial solutions are equivalent to
the twist even solution which was 
found by Takahashi and Tanimoto, and twist invariance of open string
field theory remains after the shift of the classical
backgrounds. 
\end{abstract}
\end{titlepage}

%%%%%%%%%%%%%%%%%%%%%%%%%%%%%%%%%%%%%%%%%%%%%%%%%%%%%%
\renewcommand{\thepage}{\arabic{page}}
\renewcommand{\thefootnote}{\arabic{footnote}}
\setcounter{page}{1}
\setcounter{footnote}{0}
\baselineskip=19pt plus 0.2pt minus 0.1pt
%%%%%%%%%%%%%%%%%%%%%%%%%%%%%%%%%%%%%%%%%%%%%%%%%%%%%%
%
%%%%%%%%%%%%%%%%%%%%%%%%%%%%%%%%%%%%%%%%%%%%%%%%%%%%%%

\section{Introduction}

Nowadays, string field theory (SFT) turns out to be 
one of promising candidates for the nonperturbative 
formulation of string theory. Attractive features 
of SFT --- such as background independence and local gauge symmetry
--- enable us to explore the moduli space and classical (or quantum)
dynamics of string theory.
In particular, the dynamics of unstable D-brane systems have been
extensively investigated in terms of open SFT \cite{tz-review, reviews} 
and all results obtained up to now support 
Sen's conjecture \cite{sencond}.
Furthermore, analysis of several
systems involving closed string tachyons also
have been developing \cite{closedSFT}.

In recent few years, a class of analytic classical solutions 
of cubic SFT (CSFT) \cite{csft} have been
extensively investigated \cite{TT}-\cite{super}. 
The solutions --- {\it universal solutions} \cite{TT} --- are
Lorentz invariant, universal\footnote{This means that 
solutions are independent of details of a 
boundary CFT  used there \cite{universality}.
} and conjectured to be candidates for the closed string
vacuum where there are no open strings. 
Each solution is labeled by  a function $g(w)$ 
which obeys (i) BPZ invariance, (ii) midpoint constraint and 
(iii) hermiticity. Therefore, classification of
the solutions reduce to a problem of classification of
the functions satisfying above three conditions. 

A key feature of the moduli space of the solutions
is distributions of zeros of $g(w)$ on the unit disk.
In general, it is expected that when zeros are
on the unit circle, a solution becomes
nontrivial, while when all of them are inside unit disk,
a solution is trivial and pure gauge. Such feature of
zeros was confirmed for particular cases \cite{TT, KT, fourth}
and proved for the general case of even polynomial functions
in a systematic way \cite{quad}.  

After removing trivial solutions
from the space of $g(w)$, there remain a lot of 
nontrivial solutions. If all of them correspond to unique
closed string vacuum, they must be equivalent each other
through gauge transformations. In fact, there are strong 
evidences \cite{moduli} that 
nontrivial solutions considered in \cite{KT} are equivalent 
through the `global' part of the local gauge symmetry 
generated by $K_{2m} = L_{2m} -L_{-2m}$. Such result 
allows us to interpret tachyon condensation as a dynamical
symmetry breaking concerned with the CSFT potential. 

From these results, it is natural to ask whether
classification of the whole moduli space of universal solutions
is possible. 
As mentioned before, the subspace of solutions
labeled by even polynomial functions can be analyzed 
systematically \cite{quad}. On the other hand, 
a case involving an odd part of $g(w)$ is more complicated
than the even case owing to the fact that the midpoint
condition $g(\pm i)=1$ becomes nontrivial in such case.
For such reason, the odd part have never been considered
earlier. 

Besides the problem of classification of the solution mentioned above, 
introducing an odd part of $g(w)$ poses some questions to us.
A decomposition of $g(w)$ according to worldsheet parity is
given by 
\begin{equation}
 g(w) = g_{+} (w) + g_{-} (w) ,\label{gdecomposeintro} 
\end{equation}
where $g_{\pm} (w)$ are even and odd parts of $g(w)$
which satisfy $g_{\pm} (-w) = \pm g_{\pm} (w)$ 
respectively. An important fact is that (\ref{gdecomposeintro})
yields a decomposition of the corresponding classical solution
\begin{equation}
 \Psi = \Psi_{+} + \Psi_{-}, 
\end{equation}
where $\Psi_{+}$ and  $\Psi_{-}$ are {\it twist}\footnote{
The twist operation, which is denoted as $\Omega$
usually, reverses an orientation
of an open string according to $\sigma \rightarrow \pi -\sigma$.
} even and 
odd parts of the universal solution associated with $g(w)$, respectively. 
Thus, inclusion of an odd part of $g(w)$ yields an odd component
of a classical solution.
Such case have never been considered in any attempts to
find the closed string vacuum --- level truncation in Siegel
gauge \cite{sz-tachyon, GR, mol-taylor}, vacuum string
field theory \cite{vsft} and universal solutions
\cite{TT}-\cite{moduli}. This is because of the
fact that the equation of motion with respect to $\Psi_-$ 
 is linear in $\Psi_-$ owing to twist invariance of
the CSFT action, and one can make a consistent 
truncation by setting $\Psi_-$ zero \cite{sz-tachyon}. 
However, at least in general, it is an nontrivial question
whether an nonzero twist odd part is arrowed.
If a twist odd component cannot be removed 
and has physical effects, 
such situation indicates existence of new classical vacuum 
without twist symmetry.
Of course, such vacuum contradicts the uniqueness
of the closed string vacuum, which is a part of Sen's conjecture.

Thus our questions are as follows:  twist symmetry of CSFT can be
 violated by a twist odd part of the universal solutions ? 
Does such vacuum exists ? Is it different from the closed string vacuum ?
In this paper, we would like to answer such questions 
by considering two parameter family of the solution
defined by
\begin{equation}
g(w) = 1 + \frac{a}{2}\left(w+ \frac{1}{w}\right)^2
   -i b \left(w-\frac{1}{w} + w^3 - \frac{1}{w^3} \right). 
\label{gparamintro} 
\end{equation}
The answer is that all things expected in above questions
{\it never occur}, since we can show that
every solution defined in terms
of (\ref{gparamintro}) is equivalent to the twist even solution
considered in \cite{TT} through a certain field redefinition. 
Thus whole story is consistent with
twist even solutions obtained earlier, and one does not need
to worry about violation of twist symmetry.  

The paper is organized as follows. In Section \ref{twistreviewsec},
after reviewing twist symmetry in CSFT, 
we introduce two parameter family of 
universal solutions which will be used throughout this paper.
In particular, distribution of zeros of $g(w)$ are investigated precisely. 
In following two sections, we give some evidences that the
solutions can become nontrivial when some zeros reach unit circle.
Section \ref{seccohom} deals with singularity of the field redefinition
operator and BRST cohomology of the modified BRST operator. In section
\ref{sec:level}, we give level truncation analysis of the CSFT
expanded around the classical solutions. 
In section \ref{sectwistproof} we give a conformal transformation
which completely maps our solutions to the twist even solution
considered in \cite{TT}. With the help of this map, twist symmetry of 
a CSFT around our solution can be easily understood and turns
out to be unbroken. Finally in section \ref{secconclu}, we 
summarize our results and give some discussions.

\section{Universal solution with twist odd modes}
\label{twistreviewsec}

\subsection{Twist symmetry in CSFT}

First let us summarize some aspects of CSFT concerned with 
the twist symmetry of open string theory. 
We shall borrow most notations and discussions from 
\cite{tz-review}. The CSFT action \cite{csft} on single unstable D-25 brane is given by
\begin{equation}
S[\Psi] = - \frac{1}{g_{o}^{2}} \left(
 \frac{1}{2} \braket{\Psi,  Q_B \Psi} 
+ \frac{1}{3} \braket{\Psi, \Psi, \Psi }  
\right),   \label{sftaction}
\end{equation}
where the braket $\left< \cdots \right>$ is the multi string 
product defined by the
noncommutative star products \cite{tz-review}. 
Open string theory has the twist symmetry $\Omega$ which reverses an
orientation of an open string worldsheet. The action of twist 
symmetry on the open string Hilbert space
is quite simple in the oscillator formalism.
For an oscillator satisfying $[L_0, \phi_n] = n \phi_n$, 
 \begin{equation}
  \Omega (\phi_n) = (-1)^{n} \phi_{n}.
 \end{equation}
Note that the $\mathrm{SL}(2,\mathbb{C})$ vacuum $\ket{0}$ has 
twist eigenvalue $-1$ as discussed in
\cite{tz-review}. 
Therefore we have, for example, 
\begin{equation}
  \Omega (\phi_{n} \ket{0}) = (-1)^{n+1} \phi_{n}\ket{0}.\label{twistonbase}
\end{equation}
Since the string field can be represented by
a linear combination of fock space states, (\ref{twistonbase}) 
defines the action of $\Omega$ on the string field.   
In the language of CSFT, the twist symmetry of correlation functions
of CFT is conveniently expressed as \cite{tz-review}
\begin{align}
 \braket{A, B}  & = \braket{\Omega (A), \Omega (B)}, \label{twiston2} \\
 \braket{A, B, C}  & =
 (-1)^{AB+BC+CA+1} \braket{\Omega (C), \Omega (B), \Omega (A)},
\label{twiston3}
\end{align}
where $A$, $B$ and $C$ are arbitrary string fields. 
Applying (\ref{twiston2}) and (\ref{twiston3}) to (\ref{sftaction}), 
and taking into account the facts that $\Psi$ is Grassmann odd
 and $[Q_B, \Omega ]=0$,
we can show  the twist invariance of CSFT action, 
\begin{equation}
 S[\Omega (\Psi)] = S[\Psi]. \label{twistsymmetry}
\end{equation}
In the following, we only consider a case in which 
$\Psi$ belongs to the universal subspace \cite{sz-tachyon} which 
consists of the matter Virasoro generators, conformal ghost oscillators
and the $\mathrm{SL}(2, \mathbb{C})$ vacuum. Decompose $\Psi$ as 
\begin{equation}
 \Psi = \Psi_{+} + \Psi_{-}, \label{twistdecompose}  
\end{equation}
where $\Psi_{+}$ and  $\Psi_{-}$ are string fields with twist eigenvalues
$+1$ and $-1$, respectively. Plugging (\ref{twistdecompose}) into
(\ref{sftaction}) and using (\ref{twistsymmetry}), we obtain
\begin{multline}
S[\Psi_{+} + \Psi_{-} ] = \frac{1}{2} \braket{\Psi_{+},  Q_B \Psi_{+}} 
  + \frac{1}{3} 
\braket{\Psi_{+}, \Psi_{+}, \Psi_{+} } \\
+ \frac{1}{2} \braket{\Psi_{-},  Q_B \Psi_{-}}  
 + \braket{\Psi_{+}, \Psi_{-}, \Psi_{-} }. 
\end{multline}
Equations of motion with respect to $\Psi_{+}$ and $\Psi_{-}$ are
\begin{align}
 Q_B \Psi_{+} + \Psi_{+} * \Psi_{+} + \Psi_{-} * \Psi_{-} & =0,\label{eveneom} \\
 Q_B \Psi_{-} + \Psi_{+} * \Psi_{-} + \Psi_{-} * \Psi_{+} & =0, \label{oddeom} 
\end{align}
respectively. Since (\ref{oddeom}) is linear in $\Psi_{-}$, one can 
make a consistent truncation to set $\Psi_{-}$ zero as proposed in
\cite{sz-tachyon}.  However one can also find a solution with
nonzero $\Psi_{-}$. In the following, we would like to argue such
case.

\subsection{Universal solutions}

In the formalism of universal solution \cite{TT}, which is exact
classical solutions of CSFT outside Siegel gauge 
\cite{moduli}, it is easy 
to construct solutions with twist odd string fields. 
The solution is labeled by a function $F(w)$. 
\begin{equation}
 \Psi_0  = \left\{ Q_{L} (F) - C_{L} \left(
\frac{(\partial F)^2}{1+F}
     \right) \right\} \id,
\label{solform} 
\end{equation}
where $\id$ is the identity string field. $Q_L (F)$ and
$C_{L} ((\partial F)^2/(1+F)) $ 
are integrals of the BRST current and 
conformal ghost $c(w)$ multiplied by each functions over the 
left half of an open string.  
To satisfy the equation of motion, the function $F(w)$ must obey
$F(-1/w)=F(w)$, $F(\pm i) =0$, and the hermiticity condition
which will be discussed later. 
Decompose $F(w)$ as
\begin{equation}
 F(w) = F_{+} (w) + F_{-} (w),
\end{equation}
where $F_{+} (w)$ and $F_{-} (w)$ are
even and odd functions of $w$ respectively.
Using (\ref{solform}), one can easily confirm that
setting $F_{-} (w) \neq 0$ adds twist odd components
to (\ref{solform}). 

In the rest part of this paper, 
we shall analyze the modified BRST operator obtained by
fluctuating a string field around the solution (\ref{solform}).
Such operator is defined by $Q_g \Psi = Q_B \Psi + \Psi * \Psi_0
 + \Psi_0 *\Psi$, and explicitly given by
\begin{equation}
 Q_{g} = Q(g) - C \left( \frac{(\partial g)^2}{g} \right),\label{brsdef}
\end{equation}
where $g (w) = F(w) +1$. $Q(g)$ and $C( (\partial g)^2 / g )$ 
are defined in the same manner as  $Q_{L}$ and  $C_{L}$, 
but integrals are now taken over the whole unit circle.
Again, the odd part of $g(w)$ corresponds to the contribution
of twist odd components of the solution. 

In terms of $g(w)$, the conditions imposed on $F(w)$ 
are expressed as \cite{quad}, 
\begin{equation}
 g\left(-\frac{1}{w}\right) = g(w),\label{idg} 
\end{equation}
\begin{equation}
 g(\pm i) = 1, \label{midg}
\end{equation}
\begin{equation}
 g(e^{i\theta}) \geq 0. \label{hermg}
\end{equation}
In particular, the third condition --- 
positivity of $g(w)$ on the unit circle --- 
ensures hermiticity of the modified BRST operator.
   
\subsection{Solutions with third order polynomial}

As a simple example of a solution including twist odd components, 
we shall consider
\begin{equation}
g(w) = 1 + \frac{a}{2}\left(w+ \frac{1}{w}\right)^2
   -i b \left(w-\frac{1}{w} + w^3 - \frac{1}{w^3} \right), 
\label{g2parm}
\end{equation}
where $a$ and $b$ are real parameters. With this choice,
two of the three conditions imposed on $g(w)$,
(\ref{idg}) and  (\ref{midg}), are trivially satisfied.
On the other hand, the condition (\ref{hermg}) 
will further restrict parameters $a$ and $b$ to be in certain region. 
To impose (\ref{hermg}) on (\ref{g2parm}), another parameterization 
of $g (w)$ in term of its zeros is quite useful. Such parameterization
also have been played 
an important role in the analysis of the universal 
solutions \cite{TT, KT, quad, fourth}. 
Among six zeros of (\ref{g2parm}), three zeros are
always inside the unit disk, 
while the rest are outside owing to the symmetry (\ref{idg}). 
When (\ref{hermg}) is satisfied, one of three zeros inside the
unit disk becomes pure imaginary, and other two are 
symmetric under $z \leftrightarrow -\bar{z}$. We denote the pure imaginary zero as $i t$, where $t$ is real,
and the pair of complex zeros as $x$ and $-\bar{x}$.
In terms of these zeros, we can rewrite (\ref{g2parm}) as 
\begin{equation}
 g (w) = g_{x} (w) \cdot g_{-\bar{x}} (w) \cdot 
g_{it} (w),
\label{gbyzeros}
\end{equation}
where
\begin{equation}
 g_{y} (w) = \frac{i}{(i-y)(i+y^{-1})}
\frac{(w-y)(w+y^{-1})}{w}.  \label{gxdef} 
\end{equation} 
It can be easily seen  that (\ref{gbyzeros}) satisfies 
(\ref{idg}),  (\ref{hermg}) and $g(i)=1$,
but does {\it not} obeys  $g(-i)=1$ in general. 
Of course, the last condition is necessary\footnote{
It reduces the number of independent parameters
in (\ref{gbyzeros}) to two, which is the same number of parameters 
contained in (\ref{g2parm}).}. 
Imposing $g(-i)=1$ on (\ref{gbyzeros}) gives
\begin{equation}
 \left|
x + \frac{i}{t}
\right|
= \sqrt{\frac{1}{t^2}  -1  }.\label{circleconst}
\end{equation}
This constraint means that $x$ is on a circle with radius 
$\sqrt{1/t^2 -1 }$ and center $z= -i/t$. 
Such $x$ can be parameterized as
\begin{equation}
 x\left(t, \theta \right) = - \frac{i}{t} + \sqrt{\frac{1}{t^2}  -1  }
  e^{i \theta}. \label{xparamintth}
\end{equation}
Imposing $|x|\leq 1$ on (\ref{xparamintth}) gives
\begin{equation}
 \left|\cos \theta \right| \leq \left|t \right|.
\end{equation}
For simplicity, consider a case where $t$ is positive\footnote{Negative
$t$ case can be obtained by reversing all zeros in
positive case by $z\rightarrow \bar{z}$. 
}.
For fixed $t$, the complex zeros moves inside
lower half of the unit disk along the
circle defined by (\ref{xparamintth}).  
In particular,
\begin{itemize}
 \item We exclude $t=1$ case, because in this case the 
       Laurent coefficients
       of (\ref{gbyzeros}) diverge, therefore 
       $g(w)$ no longer gives an well-defined BRST operator.
 \item The complex zeros reach to the unit circle 
       when $\cos \theta = t$. On the other hand, 
       the pure imaginary zero is always inside the
       unit disk because $0 \leq t < 1$.
\end{itemize}
Thus, the distribution of zeros is completely specified
by the two parameter family of pure imaginary and complex zeros
given by (\ref{xparamintth}) 
within the range of parameters $0 \leq t < 1 $ and $0 \leq \cos \theta \leq t$.

\section{Field redefinition and cohomology}
\label{seccohom}

Our conjecture in this and next section is 
that the classical solution given by 
(\ref{gbyzeros}) becomes nontrivial if and only if
the complex zeros ($x$ and $-\bar{x}$) reach to 
the unit circle\footnote{
It is well known that similar results for 
twist even cases hold \cite{TT, quad, fourth}.}. 
Using analytic methods, we will give two 
evidences for our conjecture. Further evidence from 
numerical analysis will be given in Sec.\ \ref{sec:level}.

\subsection{Singularity of the field redefinition} \label{sec:singularity}

First evidence for our conjecture
can be obtained by an analysis 
of field redefinitions associated with the modified BRST operator.
In the same manner as in \cite{TT, KT, quad, fourth}, 
the modified BRST operator can be formally transformed to
$Q_B$ as 
\begin{equation}
 Q_g = e^{q(h)} Q_B e^{-q(h)}.\label{qbtrans}
\end{equation}
where $h(w)=\log g(w)$ and  $q(h)$ is defined by
\begin{equation}
 q(h) = \oint dw\, h(w) j_{gh} (w),
\end{equation}
where $j_{gh} (w)$ is the ghost number current. When the operator 
$e^{q(h)}$ is regular, (\ref{qbtrans}) defines well-defined 
transformation; we can transform the CSFT action with
$Q_g$ into the original action with $Q_B$ by a field redefinition
$\Psi \rightarrow e^{-q(h)} \Psi$. In such case, a classical solution
has no physical meaning and corresponds to gauge degree of freedom. 
On the other hand, if $e^{q(h)}$ happens 
to be singular, we cannot perform such field redefinition. In such case
a solution is expected to be nontrivial. 

In order to evaluate such singularity, we shall use 
an oscillator expansion of $q(h)$. We can obtain this 
using (\ref{gbyzeros}) and
(\ref{gxdef}) and the fact that $|x| \leq 1$ and $|t| < 1$. 
The result is
\begin{equation}
q(h) = -2 \left\{\log \left(|x-i|^2 (1-t) \right)  \right\} q_{0}  -
\sum_{n=1}^{\infty}
\frac{\mathcal{A}_n (t) + \mathcal{B}_n (x)   }{n}
\left(q_{n} + (-1)^{n} q_{-n}\right),  \label{qhdef}
\end{equation}
where 
\begin{align}
 \mathcal{A}_n (t) & = (- i t)^n , \\
 \mathcal{B}_n (x) & = (-x)^n + \bar{x}^n,
\end{align}
and $q_{n}$ is the oscillator mode of $j_{gh} (w)$.
Using (\ref{qhdef}), we can evaluate the singularity of $e^{q(h)}$
by performing normal ordering with respect to the $\mathrm{SL}(2, \mathbb{C})$
vacuum. A calculation is easily performed
by using $[q_n, q_m]=n \delta_{n+m}$, and it turns out that
a contraction of two zeros $y$ and $z$ contributes a factor
$(1-yz)^{-1/2}$. A potentially divergent factor
comes from a contraction between $x$ and $-\bar{x}$ parts in $q(h)$
and it amounts to
\begin{equation}
 (1-|x|^2)^{-1}.
\end{equation}
In fact, it diverges when $x$ reaches to the unit circle.
Thus, in this case, the classical solution is expected to
be nontrivial, since CSFT action with $Q_g$ no longer equivalent to
the original action.

\subsection{Cohomology} \label{sec:cohom}

As a further evidence for our conjecture, 
we shall show that the cohomology of the modified BRST
operator vanishes when complex zeros come to unit circle. 
This means no physical excitation of open string around the
solution. Such result is expected according to
to Sen's conjecture, if our nontrivial solutions correspond
to the closed string vacuum.
The proof can be done in a way similar to \cite{KT, fourth}, 
where the cohomology of the BRST operator
is mapped to the modified Kato-Ogawa cohomology \cite{kato-ogawa} 
with `wrong' ghost numbers. 

For a critical configuration satisfying
 $|x_c|=1$ and $|t_c| < 1$,  (\ref{gbyzeros}) becomes
\begin{equation}
 g_{x_c, t_c} (w) = \frac{- i t_c}{|x_c-i|^4 (1-t_c)^2  } 
  \frac{(w-x_c)^2
(w+\bar{x}_c)^2
(w-i t_c)(w-i t_{c}^{-1})}{w^3}, \label{gcritical}
\end{equation} 
A crucial step is rewriting (\ref{gcritical}) to
\begin{equation}
g_{x_c, t_c} (w)  =  
\frac{1}{|x_c -i|^4 (1-t_c)^2} \times w^2 \times 
e^{ h_{x_c} (w) }\times  e^{h_{t_c} (w)},\label{qfac}  
\end{equation}
where
\begin{align}
 h_{x_c} (w) & = -2 \sum_{ n=1}^{\infty} 
 \left\{ 
\frac{x_{c}^{n} + (- \bar{x}_{c})^{n}   }{n} \right\} w^{-n}, \\
 h_{t_c} (w) & = - \sum_{ n=1}^{\infty} 
    \frac{(-i t_{c})^n }{n} \left( w^{n} +(-1)^{n} w^{-n} \right). 
\end{align}
Next, using the formula $Q_{e^{h} g} = e^{q(h)} Q_g  e^{-q(h)}$ \cite{TT}, we can
rewrite (\ref{qfac}) into corresponding operator equation,
\begin{equation}
 Q_g =  \frac{1}{|x_c -i|^4 (1-t_c)^2} 
e^{q(h_{x_c})}e^{q(h_{t_c})}
  Q_{B}^{(2)} e^{-q(h_{x_c})} e^{-q(h_{t_c})},\label{cohotrans}  
\end{equation}
where $Q_{B}^{(2)} = Q_{2} -2 c_{2}$ is the `shifted' BRST operator
obtained by applying a
replacement $c_n \rightarrow c_{n+2}, b_{n} \rightarrow b_{n-2}$
to $Q_B$ \cite{KT}. Note that the transformations 
in (\ref{cohotrans}) is regular, because $|t_c|< 1$  and 
$q(h_{x_c})$ consists of only negative frequency modes of 
the ghost number current.
The cohomology of $Q_{B}^{(2)}$
obtained in  \cite{KT} mapped into that of $Q_g$ as 
\begin{equation}
\ket{\Psi_{\text{phys}}} = e^{q(h_{x_c})}e^{q(h_{t_c})} 
 \left(
\ket{P} \otimes b_{-2} \ket{0} + \ket{P'} \otimes \ket{0} 
\right),
\end{equation}
where $\ket{P}$ and $\ket{P'}$ are the DDF states \cite{ddf}. Since a ghost number
of this state is not equal to unity, the state is actually 
zero in the context of classical CSFT without gauge fixing\footnote{
In this theory, only ghost number one fields are allowed. 
}.  Thus we have obtained a vanishing cohomology. 

One the other hand, for trivial solutions whose all zeros are
inside the unit disk, it is easily shown that 
the BRST cohomology is equivalent to the Kato-Ogawa cohomology\footnote{
This can be shown by mapping the Kato-Ogawa cohomology by
 $e^{q(h)}$ which appearers in (\ref{qbtrans}). }.
Therefore, our result implies that the solutions
with complex zeros on the unit circle are nontrivial 
one corresponding to the closed string vacuum.

\section{Level truncation analysis} \label{sec:level}

As another nontrivial test for our proposal, we can perform 
the level truncation analysis of the CSFT around the classical solutions.
An analysis can be done in almost same manner as \cite{fourth,
taka-level}, except for the fact that
we must include string fields of odd levels. 
In Siegel gauge, the `normalized' CSFT potential 
around the classical solution becomes 
\begin{equation}
 f_g (\Psi) = 2 \pi^2
\left(
\frac{1}{2} \braket{\Psi, c_{0} L_g \Psi} 
+ \frac{1}{3} \braket{\Psi, \Psi, \Psi }  
\right). \label{univpot} 
\end{equation}
For the representation of $g(w)$ given in (\ref{g2parm}), 
the gauge fixed kinetic operator
$L_g$ is given by 
\begin{align}
 L_g & =\left\{Q_{g}, b_{0} \right\} \notag \\
      & =(1+a)L_{0} +\frac{a}{2} \left(L'_{2} + L'_{-2} \right) 
    - i b \left( L'_{1} - L'_{-1} + L'_{3}-L'_{-3}
\right) -  \kappa (g),\label{kinticop}
\end{align}
where $L'_n = L_n + n q_n +\delta_{n,0}$ is the $c=24$ twisted 
Virasoro generator and $\kappa (g)$ is a real valued integral
defined by
\begin{equation}
 \kappa (g) = \oint
\frac{dw}{2 \pi i}  \frac{(g'(w))^2}{g(w)} \times w.
\end{equation}

Following \cite{taka-level,fourth}, let us consider a
local minimum of the potential. 
As explained in \cite{taka-level}, if our solution correctly
reproduce the D-brane tension and obeys Sen's conjecture, 
then the value of the potential at local minimum should 
depends on the positions of complex zeros as 
\begin{equation}
  f_g (\Psi_0) =
 \begin{cases}
  -1 & ( |x| < 1), \label{delta} \\
  0  &  (|x| = 1 ),
 \end{cases}
\end{equation}
where $\Psi_0$ is the local minimum. 
According to (\ref{delta}), a plot of $f_g (\Psi_0)$ with respect to parameters of $g(w)$ is expected to have 
discontinuities when $|x|=1$.
Of course, an actual plot obtained by level truncation
will never coincide with (\ref{delta}) exactly,
though it will approach (\ref{delta}) as level increases. 

The only difference between our analysis and \cite{fourth, taka-level}
is inclusion of odd level fields. 
This is necessary for our case since the 
potential (\ref{univpot}) 
is no longer invariant under $\Omega$ because
$L_g$ does not commutes with $\Omega$ in general.  
The gauge fixed equations of motion obtained from
(\ref{univpot}) become
\begin{align}
 c_{0} L_{g, +} \Psi_{+} + c_{0} L_{g, -} \Psi_{-} + 
\Psi_{+} *\Psi_{+} + \Psi_{-} * \Psi_{-} &
 =0,\label{leveleomeven} \\
 c_{0} L_{g, +} \Psi_{-} + c_{0} L_{g, -} \Psi_{+} +
 \Psi_{+} *\Psi_{-} + \Psi_{-} * \Psi_{+} &
 =0,\label{leveleomodd}  
\end{align}
where $L_{g, +}$ and $L_{g, -}$ are twist even and odd parts of
(\ref{kinticop}), respectively. Note that $\Psi_{-}$ cannot be
set to zero to satisfy (\ref{leveleomodd})  
because of the presence of the odd part of the kinetic operator.

Now we shall try to find a level truncated solution of (\ref{leveleomeven}) and
(\ref{leveleomodd}) in the universal subspace \cite{sz-tachyon}.
Actual calculation is done as follows:
\begin{itemize}
 \item  Required techniques to perform level truncation analysis 
	have already developed by many authors, and they can also be
	applied for our case. To compute three string vertex, we use
	the conservation current method and an algorithm used
        in \cite{GR}. 

 \item  We can easily see that there are no level 1 fields in Siegel
	gauge. Odd fields appearers first at level 3. That
	is given by, 
	\begin{equation}
	 \ket{\Psi_{\text{level } 3}} = 
	  i s_1 c_{-2} \ket{0} +i s_2 b_{-2}c_{-1}c_{1} \ket{0}
     + i s_3 L^{(X)}_{-3}c_{1} \ket{0},\label{level3}
	\end{equation}
where $s_1$, $s_2$ and $s_3$ are real parameters. 

 \item We have done all computations up to level (6, 18).  
\end{itemize}
Since our solution has two independent parameters, we must
fix one of them to draw a two dimensional plot of potential values. We will do this in two different ways.

\subsection{$a=b$ case} \label{sec:aeqb}

\begin{figure}[htb]
\centerline{\includegraphics[scale=0.6,clip]{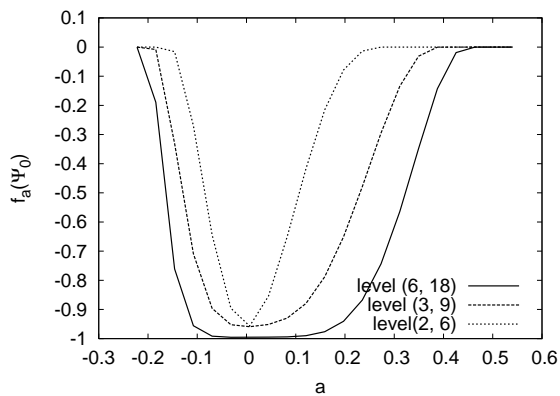}}
\caption{A plot of potential value of $a=b$ solution}
\label{potentialplot}
\end{figure}

As first example of one parameter solution, we set 
$a=b$ in (\ref{g2parm}). This gives 
\begin{equation}
 g(w) = 1 + \frac{a}{2}\left(w+ \frac{1}{w}\right)^2
   -i a \left(w-\frac{1}{w} + w^3 - \frac{1}{w^3} \right)
.\label{goriginal}
\end{equation}
The condition (\ref{hermg}) determines the range of $a$ as
\begin{equation}
   - \frac{2}{9} \leq a \leq \frac{27}{50}. \label{arange}
\end{equation}
Two boundaries of the above region correspond to critical points
where the complex 
zeros are on the unit circle. Positions of zeros at
two critical points are given by
\begin{equation}
 (x_c, i t_c) = \begin{cases}
	   \left(\frac{\sqrt{3} + i}{2}, -\frac{1}{2}i \right) & (a =
	   -\frac{2}{9}) \\
	   \left(\frac{\sqrt{5}- 2 i}{3},\frac{2}{3} i \right) & (a = -\frac{27}{50}) 
	  \end{cases}
\end{equation}
According to (\ref{delta}), the potential value
will depend on $a$ as
\begin{equation}
  f_a (\Psi_0) =
 \begin{cases}
  -1 & ( - \frac{2}{9} < a < \frac{27}{50}), \label{deltaa} \\
  0  &  (a = - \frac{2}{9}, \frac{27}{50}).
 \end{cases}
\end{equation}
Using (\ref{goriginal}), we can compute the value of the
potential at a local minimum $f_a (\Psi_0)$ as a function of $a$. 
Figure \ref{potentialplot} is a plot of our result up to level 6. 
We can see that a flat region at
the bottom glows, and two flat regions at
both sides shrink as level increases.  
Surely curves approach the discontinuous configuration
predicted in (\ref{deltaa}).
In addition, we can see that
level 3 fields --- first odd fields ---
surely contribute to the potential value and improve
level 2 result. 

Table \ref{fieldtbl} picks up values of some fields at
a local minimum in the case of $a=0.1$. 
$t$ is the level 0 field, and $s_1$ and $s_3$ are
level 3 fields appeared in (\ref{level3}).  
Indeed, we can see that level 3 fields acquire nonzero 
values as expected. 

\begin{table}[htb]
\begin{center}
  \begin{tabular}{|c|c|c|c|c|}
  \hline
 level & $t$ & $s_3$ & $s_1$ &
 $f_a (\Psi_0)$ \\ \hline\hline
 (3, 9) & 0.5161 & -0.0168 & 0.0429 & -0.9111 \\ \hline
 (4, 12) & 0.5298 & -0.0181 &  0.05119 & -0.9812 \\ \hline
 (5, 15) & 0.5260 &   -0.0182 & 0.0541 & -0.9747\\ \hline
 (6, 18) & 0.5259 & -0.0187 & 0.0543 & -0.9928 \\ \hline
 \end{tabular}
\end{center}
\caption{Value of some string fields at $a=0.1$}
\label{fieldtbl}
\end{table}

\subsection{Fixed $t$ case} 

\begin{figure}[bht]
\centerline{\includegraphics[clip]{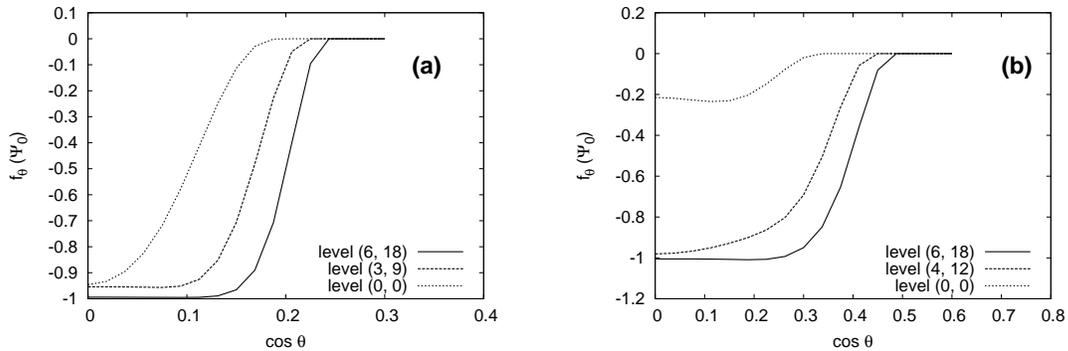}}
\caption{Plots of the potential values at (a) $t=0.3$ 
and (b) $t=0.7$ with respect to
$\cos\theta$}
\label{plot-t}
\end{figure}

Next we consider another one parameter family of 
our solutions obtained by fixing a position of pure imaginary
zero appearers in (\ref{gbyzeros}). The relation between
zeros $x$ and $t$ is already given in (\ref{xparamintth}). 
By fixing $t$ at some value $t_0$, we have
\begin{equation}
 x\left(\theta \right) = - \frac{i}{t_0} + \sqrt{\frac{1}{t_{0}^{2}}  -1
  } \, e^{i \theta}. \label{xparamintthzero}
\end{equation}
In order to $x(\theta)$ to be inside unit disk, we must restrict
$\theta$ to $ 0 \leq \cos \theta \leq t_0$. 
Thus (\ref{gbyzeros}) and (\ref{xparamintthzero}) define
one parameter family of solution for each $t_0$ with parameter $\theta$.
$x(\theta)$ reaches the unit circle
when $\cos \theta = t_0$, so this value of $\theta$ corresponds to
critical configuration where solution expected to be nontrivial. 
Again, according to (\ref{delta}), 
the parameter dependence of the potential value 
will be
\begin{equation}
  f_{\theta} (\Psi_0) =
 \begin{cases}
  -1 & ( \cos \theta < t_0),  \\
  0  &  ( \cos \theta = t_0 ).
 \end{cases} 
\end{equation}
Now, we can perform level truncation analysis for various values of $t$
in the same manner as in Sec.\ \ref{sec:aeqb}. 
Figures in \ref{plot-t} are plots of the
potential value at $t=0.3$ and $t=0.7$ respectively. Again, 
curves in the plots approach to desired one as level increases. 
In both plots, a flat region at the left bottom of the curves grows, 
and another flat region at the right top 
becomes narrower as level goes higher.

\section{Twist symmetry around classical solutions}
\label{sectwistproof}

The CSFT action expanded around our solution 
is no longer invariant under the twist operation $\Omega$ in general.
For nonzero $b$, (\ref{g2parm}) has odd part, so
$Q_g$ includes twist odd components.  Consequently, 
$\Omega$ no longer commutes with $Q_g$. From this fact, we can see
\begin{equation}
 S_g [\Omega (\Psi)] \neq  S_g [\Psi], 
\end{equation}
where $S_g[\Psi] = 1/2 \braket{\Psi, Q_g \Psi} + 1/3
\braket{\Psi, \Psi, \Psi}$. 
An naive interpretation of the above fact is that 
the twist symmetry of original CSFT is broken at the
classical vacuum specified by one of our solution.
However, at least for a trivial solution, 
such phenomenon never occur,
since a CSFT around a trivial solution is equivalent to the original CSFT 
described by $Q_B$, which has twist invariant action.  
Therefore the twist symmetry never be violated
in a CSFT around a trivial solution. This fact can be easily
confirmed by mapping the twist operator $\Omega$ to new one. 

Consider the function $g(w)$ such like (\ref{g2parm})
whose zeros are inside the
unit disk. Then the CSFT action around the solution associated
with $g(w)$ is described by the modified BRST operator $Q_g$.
In this case, one can always find a regular transformation\footnote{
For example, one can set $U_g=e^{q(\log g)}$. Also $U$
can be taken to be a conformal transformation generated by 
$K_n = L_n -(-1)^n L_{-n}$  } which satisfy 
\begin{equation}
U Q_g U^{-1} = Q_B. \label{regulartr}  
\end{equation}
Using $U$, we can construct the deformed twist operator
\begin{equation}
 \widetilde{\Omega} = U \Omega U^{-1}. \label{omegatrans}
\end{equation}
Since $U$ is regular, $\widetilde{\Omega}$ has same algebraic 
properties as the original twist. Furthermore, $\widetilde{\Omega}$ 
leaves $S_{Q_g}[\Psi]$ invariant. Therefore, twist invariance of the 
SFT with $Q_g$ is represented by $\widetilde{\Omega}$.

On the other hand, for nontrivial solutions whose some zeros 
of $g(w)$ are on the unit circle, 
a situation is quite different.
An attempt to find regular transformation like
(\ref{regulartr}) always fails.
This is because CSFT around nontrivial solution
is no longer equivalent to the original theory with $Q_B$.
Thus we cannot use (\ref{regulartr}) to obtain new
twist operator.

However, if there exists another twist even BRST operator
which can be connected smoothly to $Q_g$, similar argument
to the case of trivial solution is possible.
We would like to try to find a regular transformation and
a twist even BRST operator which satisfy
\begin{equation}
U Q_g U^{-1} = Q_{\tilde{g}}, \label{nontritrans}
\end{equation}
where $\tilde{g} (w)$ is an {\it even} function whose
some zeros are on the unit circle. Of course, $Q_{\tilde{g}}$
becomes twist even operator associated with nontrivial solution. 
Once $U$ is find, new twist operator which leaves $S_g [\Psi]$
invariant can be constructed in the same way 
as the case of trivial solutions. 

To find regular transformation in (\ref{nontritrans})
, let us consider one parameter family of nontrivial solutions whose 
the complex zeros are {\it always} on the unit circle.
Setting $\cos \theta = t$ in (\ref{xparamintth})
realize such situation. With this choice, $x$ becomes
\begin{equation}
 x = \sqrt{1-t^2} - i t. \label{xtsingular}
\end{equation}
Inserting (\ref{xtsingular}) into (\ref{gbyzeros}), 
we find 
\begin{equation}
 g_{t} (w) = 1 +\frac{a(t)}{2}\left(w+\frac{1}{w}\right)^2
-i b(t) \left(w-\frac{1}{w}+w^3 - \frac{1}{w^3}\right) \label{gcriticals}
\end{equation}
where
\begin{equation}
a(t) = \frac{3t^2-1}{2(1-t^2)^2}, \quad
b(t) = \frac{t}{4(1-t^2)^2}.  \label{critab}
\end{equation}
Note that $b(t)$ vanishes at $t=0$. 
This means that the BRST operator becomes twist even
at $t=0$. In fact, this case is nothing but the
nontrivial Takahashi-Tanimoto solution \cite{TT}
which is given by
\begin{equation}
g_{TT} (w) = g_{0} (w) = -\frac{1}{4} \left(w+\frac{1}{w}\right)^2.
\end{equation} 

Following (\ref{nontritrans}), let us try to find $U(t)$ satisfying
\begin{equation}
 U(t) Q_{t} U(t)^{-1} = Q_{0}, \label{Qtrelation}
\end{equation}
where $Q_t$ denotes $Q_{g_t}$.\footnote{Note that
$Q_{0}$ is the nontrivial Takahashi-Tanimoto BRST operator.} 
It turns out that an attempt to find $U(t)$ 
in terms of ghost current transformation fails. Thus we
try to find $U(t)$ in terms of conformal transformation. 
By a formal argument of contour integrals of the BRST currents,
(\ref{Qtrelation}) can be reduced to the equation
\begin{equation}
 g_{t} (f_{t}^{-1} (w)) = g_{0} (w),
\end{equation}
where $f_{t} (w)$ is the finite conformal map generated by $U(t)$.
Though this equation can be solved, it turns out that
the generator of $U(t)$ becomes an infinite sum of
Virasoro generators $K_n= L_n -(-1)^n L_{-n}$. This makes it difficult
to see whether $U(t)$ is regular. Instead of dealing with
this transformation, let us consider more simpler one. Our ansatz is 
\begin{equation}
 g_{t} (f_{t}^{-1} (w)) = g_{0} (w) \times \sigma(w),
\label{newansatz} 
\end{equation}
where $\sigma(w)$ has no poles or zeros on the
unit circle. With such choice, $\sigma(w)$ can be absorbed into regular
field redefinition and has no physical effect.
In order to archive (\ref{newansatz}), $f_t (w)$
must map zeros of $g_{t} (w)$ on the unit circle 
to that of $g_0 (w)$. From (\ref{xtsingular}),
this condition is explicitly represented  as
\begin{equation}
 f_{t} (x(t)) =1, \quad f_{t} (- \bar{x}(t)) = -1. \label{slcond}
\end{equation}
It turns out that $\mathrm{SL}(2,\mathbb{C})$ is enough to satisfy the above
condition. From (\ref{xtsingular}), we can obtain 
such conformal map satisfying (\ref{slcond}):
\begin{equation}
 f_t (w)=
\frac{A(t) w + B(t)}{-B(t)w+  A(t)}, \label{sl}
\end{equation}
where
\begin{equation}
 \quad A(t) = \cosh \left(\frac{1}{2}\tanh^{-1} t\right),
\quad B(t) = i \sinh \left(\frac{1}{2}\tanh^{-1} t\right).
\end{equation}
Finally, using the explicit expression (\ref{gcriticals}) and
(\ref{critab}), and plugging (\ref{sl}) into (\ref{newansatz}), 
we have
\begin{equation}
 \sigma (w) =-4w^2 \frac{
  \left\{
(t^3 +3t)w^2 -2i(3 t^2 +1)w -(t^3+ 3t)
\right\}
}
{
\left(
tw^2 -2iw -t
\right)^3
}.
\end{equation}
Indeed, it is easily confirmed that $\sigma(w)$ has
no poles or zeros on the unit circle when $0 \leq t < 1$.
Now that $\sigma (w)$ turns out to give regular field redefinition,
the decomposition (\ref{newansatz}) can be rewritten into
operator equation. This gives our final formula 
\begin{equation}
 U(t) Q_t U(t)^{-1} = e^{q(\Sigma)} Q_0 e^{-q(\Sigma)},
\end{equation}
where $\Sigma(w) = \log \sigma (w)$ and
\begin{equation}
 U(t) = \exp\left\{
\frac{i}{2}\left(\tanh^{-1}t \right) K_1  \right\}.
\end{equation} 
As desired, both $U(t)$ and $e^{q(\Sigma)}$ are 
regular transformations. Therefore, $Q_t$ is
transformed into twist even operator $Q_0$ 
by a regular transformation. New twist operator which 
leaves $S_{g_t}[\Psi]$ invariant is
\begin{equation}
 \Omega' = e^{-q(\Sigma)} U(t) \Omega U(t)^{-1} e^{q(\Sigma)}.
\end{equation}
Thus we have shown that nontrivial solutions
with twist odd modes considered in this paper is equivalent to the 
twist even nontrivial solution specified by $g_{0} (w)$.

\section{Conclusions and discussions}
\label{secconclu}

In this paper, we have constructed a simple example of 
classical solutions of CSFT which include twist odd modes. 
A detailed analysis of the meromorphic function which governs
the solution shows that our solutions become nontrivial when
complex zeros of the function reach the unit circle. 

Furthermore, it was shown that 
the BRST cohomology of the kinetic operator of CSFT expanded 
around the nontrivial solution
is equivalent to that of the twist even solution given in \cite{TT}.
Also we have performed level truncation analysis using the CSFT potential
obtained by expanding original CSFT action around our solutions.
Our plots of the value of the CSFT potential 
with respect to positions of the complex zeros indicate that our solutions
are consistent with Sen's conjecture and nontrivial solutions
correspond to the closed string vacuum.

Finally, we have shown that our nontrivial solutions are equivalent
to the twist even solutions found in \cite{TT} by constructing a
conformal map which connects BRST operators associated with 
each solution. With the help of this map, the twist operator 
of CSFT expanded around our solutions was obtained 
from ordinary twist operator $\Omega$. Thus, our result suggests that
the twist symmetry is unbroken even for our solutions which contain
twist odd modes, and supports the uniqueness of the closed string
vacuum. 

While our analysis is done in the classical CSFT where no gauge
fixing condition is imposed, it is interesting to consider same
situation as ours, i.e., classical solutions involving twist odd
modes in the CSFT in Siegel gauge. We have found that there are no
twist odd excitations which reproduce the value of the potential
close to the D-25 brane tension, at least up to level 3. 
It would be nice if there is general arguments whether nonzero 
values of twist odd fields are possible.

It is also interesting to consider more general solutions than the
case considered in this paper. 
If we limit ourself to the polynomial case,
a typical form of the function $g(w)$ will be
\begin{equation}
 g(w) = \sum_{ n=0}^{N} a_n (w^n   + (-1)^n w^{-n}). \label{generalansatz}
\end{equation}
The most crucial point is to specify the distribution of
zeros under three conditions imposed on $g(w)$.  If this becomes clear,
it is straightforward to classify all solutions in this  class.

Our results and earlier works 
\cite{TT}-\cite{super} are  enough to expect that all
nontrivial solutions belong to the class of (\ref{generalansatz}) 
are equivalent, thus correspond to the unique solution i.e.,  closed string
vacuum. Such consideration of the moduli space of SFT
will give us  deep insights into the nature of SFT and string theory.

%%%%%%%%%%%%%%%%%%%%%%%%%%%%%%%%%%%%%%%%%%%%%%%%%%%%%%%%%%%%%%%%
\section*{Acknowledgement}

The author would like to thank I. Kishimoto, S. Moriyama, 
H. Shimada, S. Sugimoto,
T. Takahashi, S. Teraguchi and T. Tokunaga  for valuable discussions 
for valuable discussions and comments. 
I also thank to Summer Institute String Theory 2005 at Sapporo, Japan, 
for giving me an opportunity to discuss with many people. 

%%%%%%%%%%%%%%%%%%%%%%%%%%%%%%%%%%%%%%%%%%%%%%%%%%%%%%%%%%%%%%%%

%%%%%%%%%%%%%%%%%%%%%%%%%%%%%%%%%%%%%%%%%%%%%%%%%%%%%%%%%%%%%%
%\newpage

%%%%%%%%%%%%%%%%%%%%%%%%%%%%%%%%%%%%%%%%%%%%%%%%%%%%%%%%%%%%%
\newcommand{\npb}[3]{Nucl.\ Phys.\ {\bf B  #1} (#2) #3}
\newcommand{\ptp}[3]{Prog.\ Theor.\ Phys.\ {\bf #1} (#2) #3}
\newcommand{\prl}[3]{Phys.\ Rev.\ Lett.\ {\bf #1} (#2) #3}
\newcommand{\prd}[3]{Phys.\ Rev.\ {\bf D  #1} (#2) #3}
\newcommand{\plb}[3]{Phys.\ Lett.\ {\bf B #1} (#2) #3}
\newcommand{\jhep}[3]{J.\ High~Energy~Phys.\ {\bf #1} (#2) #3 }
\newcommand{\jmp}[3]{J.\ Math.\ Phys.\ {\bf #1}  (#2) #3}
\newcommand{\hepth}[1]{{\tt hep-th/#1}}
\newcommand{\ijmpa}[3]{Int.\ J.\ Mod.\ Phys.\ {\bf A #1}  (#2) #3}
\newcommand{\atmp}[3]{Adv.\ Theor.\ Math.\ Phys.\ {\bf #1}  (#2) #3}
\newcommand{\ap}[3]{Ann.\ Phys.\ {\bf #1}  (#2) #3}


\begin{thebibliography}{99}

\bibitem{tz-review} 
W.~Taylor and B.~Zwiebach,
 {\it D-Branes, tachyons, and string field theory},
\hepth{0311017}. 

\bibitem{reviews}
K.~Ohmori, {\it A Review on Tachyon Condensation in Open String Field
	Theories}, \hepth{0102085};
P.~J.~De Smet, 
 {\it A Review on Tachyon Condensation in Open String Field
	Theories}, \hepth{0109182};
I.~Ya.~Aref'eva, D~.M.~Belov, A.~A.~Giryavets, A~.S.~Koshelev
and  P.~B.~Medvedev
{\it Noncommutative Field Theories and (Super)String Field Theories},
\hepth{0111208};
L.~Bonora, C.~Maccaferri, D.~Mamone and M.~Salizzoni,
{\it Topics in String Field Theory},
\hepth{0304270}.

\bibitem{sencond}
A.~Sen, {\it Descent relations among bosonic D-branes},
\ijmpa{14}{1999}{4061} [\hepth{9902105}].

\bibitem{closedSFT}
Y.~Okawa and B.~Zwiebach,
{\it Twisted Tachyon Condensation in Closed String Field Theory},
\jhep{03}{2004}{056} [\hepth{0403051}];
H.~Yang and B.~Zwiebach,
{\it Testing Closed String Field Theory with Marginal Fields},
\jhep{06}{2005}{038} [\hepth{0501142}];
H.~Yang and B.~Zwiebach,
{\it  Dilaton Deformations in Closed String Field Theory},
\jhep{05}{2005}{032};
H.~Yang and B.~Zwiebach,
{\it Rolling Closed String Tachyons and the Big Crunch},
\hepth{0506076};
H.~Yang and B.~Zwiebach,
{\it A Closed String Tachyon Vacuum ?},
\hepth{0506077}.

%\bibitem{inner}
%G. T. Horowitz and A. Strominger, 
%{\it Translations as Inner Derivations and Associativity Anomalies in Open String Field Theory},
%\plb{185}{1987}{45}.

\bibitem{csft} 
E.~Witten,
{\it Non-commutative geometry and string field theory},
\npb{268}{1986}{253}.

\bibitem{TT}
T.~Takahashi and S.~Tanimoto,
{\it Marginal and scalar solutions in open cubic string field theory},
\jhep{03}{2002}{033} [\hepth{020133}].

\bibitem{KT}
I.~Kishimoto and T.~Takahashi,
{\it Open string field theory around universal solutions},
\ptp{108}{2002}{591} [\hepth{0205275}].

\bibitem{taka-level}
T.~Takahashi, 
{\it Tachyon condensation and universal solutions in string field theory},
\npb{670}{2003}{161} [\hepth{0302182}].

\bibitem{TZ}
T.~Takahashi and S.~Zeze,
{\it Gauge Fixing and Scattering Amplitudes in String Field Theory 
Expanded around Universal Solutions}, 
\ptp{110}{2003}{159} [\hepth{0304261}].

\bibitem{quad}
S.~Zeze, 
{\it World sheet geometry of classical solutions in string field theory}, 
\ptp{112}{2004}{863} [\hepth{0405097}].

\bibitem{fourth}
Y.~Igarashi, K.~Itoh, F.~Katsumata, T.~Takahashi and S.~Zeze,
{\it Classical solutions and order of zeros in open string field theory},
\hepth{0502042}.

\bibitem{moduli}
Y.~Igarashi, K.~Itoh, F.~Katsumata, T.~Takahashi and S.~Zeze,
{\it Exploring vacuum manifold of open string field theory},
\hepth{0506083}.

\bibitem{super}
I.~Kishimoto and  T.~Takahashi,
{\it Marginal Deformations and Classical Solutions in 
Open Superstring Field Theory},
\hepth{0506240}.

\bibitem{universality}
A.~Sen,
{\it Universality of the Tachyon Potential},
\jhep{12}{1999}{027}.

 \bibitem{sz-tachyon} A.~Sen and B.~Zwiebach,
{\it Tachyon condensation in string field theory},
\jhep{03}{2000}{002} [\hepth{9912249}].
 
\bibitem{GR}
D.~Gaiotto and L.~Rastelli,
{\it Experimental string field theory}, 
\jhep{08}{2003}{048} [\hepth{0211012}]

\bibitem{mol-taylor}
N.~Moeller and W.~Taylor,
{\it Level truncation and the tachyon in open bosonic 
string field theory}, 
\npb{538}{2000}{105} [\hepth{0002237}].

\bibitem{vsft}
 L.~Rastelli, A.~Sen and  B.~Zwiebach,
{\it String Field Theory Around the Tachyon Vacuum},
\atmp{5}{2002}{353} [\hepth{0012251}]; 
 D.~Gaiotto, L.~Rastelli, A.~Sen and  B.~Zwiebach,
{\it Ghost Structure and Closed Strings in Vacuum String Field Theory},
\atmp{6}{2003}{403} [\hepth{0111129}]. 

\bibitem{kato-ogawa}
M.~Kato and K.~Ogawa,
{\it Covariant quantization of string based on BRS invariance},
\npb{212}{1983}{443}.

\bibitem{ddf}
E.~Del~Giudice, P.~Di~Vecchia and S.~Fubini,
{\it General properties of the dual resonance model},
\ap{70}{1972}{378}

\end{thebibliography}
\end{document}